\documentclass[prd,aps,superscriptaddress,showpacs,amssymb,amsmath,twocolumn,floatfix]{revtex4}
\usepackage{bm,graphicx,dcolumn}

\newlength{\myfigwidth}
\newcommand{\Tr}{\mathop{\textrm{Tr}}}
\newcommand{\bk}{\mathbf{k}}

\newcommand{\bn}{\mathbf{n}}
\newcommand{\be}{{\bm\epsilon}}
\newcommand{\smallfrac}[2]{\genfrac{}{}{0pt}{1}{#1}{#2}}
\newcommand{\bacol}{\setlength{\arraycolsep}{0pt}}

\begin{document}

\title{Unquenching effects on the coefficients of the L\"uscher-Weisz action}

\author{Zh. \surname{Hao}}
 \affiliation{Simon Fraser University, Department of Physics,
              8888 University Drive, Burnaby, BC, V5A 1S6, Canada}
\author{G.M. \surname{von Hippel}}
 \affiliation{Department of Physics, University of Regina,
              Regina, Saskatchewan, S4S 0A2, Canada}
\author{R.R. \surname{Horgan}}
 \affiliation{DAMTP, University of Cambridge, Centre for Mathematical Sciences,
              Cambridge CB3 0WA, United Kingdom}
\author{Q.J. \surname{Mason}}
 \affiliation{DAMTP, University of Cambridge, Centre for Mathematical Sciences,
              Cambridge CB3 0WA, United Kingdom}
 \affiliation{Barclays Capital, 5 The North Colonnade,
              Canary Wharf, London E14 4BB, United Kingdom}
\author{H.D. \surname{Trottier}}
 \affiliation{Simon Fraser University, Department of Physics,
              8888 University Drive, Burnaby, BC, V5A 1S6, Canada}
 \affiliation{TRIUMF,
              4004 Westbrook Mall, Vancouver, BC, V6T 2A2, Canada}

\collaboration{HPQCD collaboration}

%% 12.38.Gc : QCD -- Lattice QCD calculations
%% 12.38.Bx : QCD -- Perturbative calculations

\preprint{DAMTP-2006-124}
%%\preprint{SFU HEP-....}
%%\preprint{TRI-PP-06-..}
%% Regina does not use preprint numbers

\begin{abstract}
  The effects of unquenching on the perturbative improvement
  coefficients in the Symanzik action are computed within the
  framework of L\"uscher-Weisz on-shell improvement. We find
  that the effects of quark loops are surprisingly large, and
  their omission may well explain the scaling violations observed in
  some unquenched studies.
\end{abstract}

\maketitle

\section{Introduction}
\label{sec:introduction}

The enormous advances in parallel computing made during the past few
years, together with theoretical advances in the formulation of
lattice gauge theories with fermions, have allowed lattice theorists
to abandon the quenched approximation that dominated lattice QCD
simulations for such a long time in favour of simulations using
dynamical light quarks. This important step has allowed a significant
reduction in systematic errors by removing the large and uncontrolled
errors inherent in the quenched approximation.

The Fermilab Lattice, MILC and HPQCD collaborations have
an ambitious program which to date has made several high-precision
predictions from unquenched lattice QCD simulations
\cite{Davies:2003ik, Aubin:2004fs},
including accurate determinations of the strong coupling constant
$\alpha_s$
\cite{Mason:2005zx},
the light and strange quark masses
\cite{Mason:2005bj},
and the leptonic and semileptonic decays of the $D$ meson
\cite{Kronfeld:2005fy}.
To do this, we rely on the Symanzik-improved staggered-quark
formalism
\cite{Lepage:1998vj},
specifically the use of the asqtad
\cite{Orginos:1999cr}
action. While this approach requires the use of the fourth root of the
staggered quark action determinant, all of the available evidence to
date is consistent with the conclusion that the resulting theory is in
the same universality class as continuum QCD, as long as the chiral
limit is taken after the continuum limit \cite{Sharpe:2006re}.

Recent studies of the heavy-quark potential in full QCD
\cite{Davies:private}
have shown an apparent increase in scaling violations compared to the
quenched approximation, contrary to
expectations. A possible reason for this would be that these scaling
violations arise from the mismatch between the inclusion of sea
quark effects in the simulation and the omission of sea quark effects
in the improvement coefficients in the action, which would appear to
spoil the $\mathcal{O}(a^2)$ improvement at the level of
$\mathcal{O}(\alpha_s N_f a^2)$. A systematic study of
$\mathcal{O}(\alpha_sa^2)$ effects is generally beyond the scope of
the current perturbative improvement programme.
Nevertheless, it is important to bring up-to-date the calculation by
L\"uscher and Weisz
\cite{Luscher:1985wf}
and by Snippe
\cite{Snippe:1997ru}
of the radiative correction to the $\mathcal{O}(a^2)$ tree-level
Symanzik-improved gluon action to include the effects of dynamical quarks.
This is important also because the L\"uscher-Weisz improvement is currently
included in many unquenched simulations 
\cite{Orginos:1999cr}.
Since the lattice spacing scale is set by measurement of the
heavy-quark potential, there will be an induced
$\mathcal{O}(\alpha_s N_f a^2)$ artifact by omitting the corrections
due to unquenching.
While such errors are generally smaller than other systematic errors in
current state-of-the art studies, it is simple to remove them, using the
result of the perturbative matching calculations done here, and this may
prove advantageous in careful studies of different scale setting
procedures.

In this paper, we present the determination of the lowest-order
perturbative contributions from quark loops to the Symanzik
improvement coefficients of the L\"uscher-Weisz glue action. Including
these contributions in future simulations, as well as accounting for
their influence in the analysis of existing results, should help to
eradicate the last remaining vestiges of the quenched approximation
and any associated systematic errors from unquenched lattice
results. Some of this work has been reported in preliminary form in
\cite{Hao:2006}.

\section{Concepts and Methods}
\label{sec:concepts}

First, let us briefly explain the ingredients of our calculation.

\subsection{On-shell improvement}
\label{subsec:onshell}

The original Symanzik improvement programme
\cite{Symanzik:1983dc,Symanzik:1983gh}
aims to remove the discretisation artifacts from the correlation
functions of the lattice theory. For gauge theories, this has proven
difficult to implement, since the correlation functions themselves are
not gauge invariant. A way out of this difficulty is offered by the
method of on-shell improvement introduced by L\"uscher and Weisz
\cite{Luscher:1984xn,Luscher:1985zq}
which aims to improve only gauge invariant spectral
quantities.

The L\"uscher-Weisz action is given by
\cite{Luscher:1984xn,Alford:1995hw}
\begin{eqnarray}
S & = & \sum_x \Bigg\{
c_0 \sum_{\mu\not=\nu}\left<1-P_{\mu\nu}\right>
+2 c_1 \sum_{\mu\not=\nu}\left<1-R_{\mu\nu}\right> \label{eqn:lw_action}\\
&&
+\smallfrac{4}{3} c_2 \sum_{\mu\not=\nu\not=\rho}\left<1-T_{\mu\nu\rho}\right>
\Bigg\}\;,\nonumber
\end{eqnarray}
where $P$, $R$ and $T$ are the plaquette, rectangle and ``twisted''
parallelogram loops, respectively.

The requirement of obtaining the Yang-Mills action in the continuum
limit imposes the constraint
\begin{equation}
c_0 + 8 c_1 + 8 c_2 = 1\;,
\end{equation}
which can be used to determine $c_0$ in terms of the other two
coefficients. This leaves us with $c_1$ and $c_2$ as unknown
coefficients which need to be determined in order to eliminate
the $\mathcal{O}(a^2)$ lattice artifacts.

If we have two independent quantities $Q_1$ and $Q_2$ which can be
expanded in powers of $(\mu a)$, where $\mu$ is some energy scale, as
\begin{equation}
Q_i = \bar{Q}_i + w_i (\mu a)^2 + 
\mathcal{O}\left((\mu a)^4\right)
\end{equation}
and which receive corrections
\begin{equation}
\Delta_\textrm{imp} Q_i = d_{ij} c_j (\mu a)^2 + 
\mathcal{O}\left((\mu a)^4\right)
\end{equation}
from the improvement operators, then the $\mathcal{O}(a^2)$ matching
condition reads
\begin{equation}\label{eqn:impcond_generic}
d_{ij} c_j = -w_i\;.
\end{equation}
Since this equation is linear, we can decompose the $w_i$ into a
gluonic and a fermionic part as $w_i = w_i^\textrm{glue} + N_f
w_i^\textrm{quark}$ and obtain the same decomposition for the $c_i$;
thus, especially we do not need to repeat the quenched calculation
\cite{Luscher:1985wf,Snippe:1997ru}
in order to obtain the $\mathcal{O}(N_f)$ contributions (however, doing
so provides a useful check on the correctness of our methods, which we
have performed successfully). At higher orders in perturbation theory,
the $d_{ij}$ and $w_i$ will become functions of the $c_i$ in lower orders.

At the tree-level, the fermions contribute nothing to gluonic
observables, and hence the tree-level coefficients remain unchanged
compared to the quenched case \cite{Luscher:1985wf}:
\begin{eqnarray}
c_1 & = & -\frac{1}{12}\;, \nonumber \\
c_2 & = & 0.
\end{eqnarray}

\subsection{Lattice perturbation theory}
\label{subsec:lpt}

Lattice field theory is usually employed as a non-perturbative
regularisation; for the calculations we need to perform, however, we
need a perturbative expansion of Lattice QCD.

In lattice perturbation theory, the link variables $U_\mu$ are
expressed in terms of the gauge field $A_\mu$ as
\begin{equation}
U_\mu(x) =
\exp\left(g a A_\mu\left(x+\smallfrac{1}{2}\hat{\mu}\right)\right)
\end{equation}
which, when expanded in powers of $g$, leads to a perturbative
expansion of the lattice action, from which the perturbative vertex
functions can be derived.

The gauge field $A_\mu$ is Lie algebra-valued, and can be decomposed
as
\begin{equation}
A_\mu(x) = \sum_a A_\mu^a(x) t^a\;,
\end{equation}
with the $t^a$ being anti-hermitian generators of SU($N$), where $N=3$
in the case of QCD.

As in any perturbative formulation of a gauge theory, gauge fixing and
ghost terms appear in the Fadeev-Popov Lagrangian; here we will not
have to concern ourselves with these, since for the purpose of
determining the unquenching effects at one loop we only need to
consider quark loops. An additional term, which we also do not need to
consider here, arises from the Haar measure on the gauge group.

The loop integrals of continuum perturbation theory are replaced by
finite sums over the points of the reciprocal lattice in lattice
perturbation theory, or integrals over the Brillouin zone where the
lattice has infinite spatial extent.

To handle the complicated form of the vertices and propagators in
lattice perturbation theory, we employ a number of automation methods
\cite{Drummond:2002kp,Hart:2004bd,Nobes:2001tf,Nobes:2003nc,Trottier:2003bw}
that are based on the seminal work of L\"uscher and Weisz
\cite{Luscher:1985wf}.
Three independent implementations by different authors have been
used in this work to ensure against programming errors.

\subsection{Twisted boundary conditions}
\label{subsec:twistedbc}

We work on a four-dimensional Euclidean lattice of length $La$ in the
$x$ and $y$ directions and lengths $L_za,~L_ta$ in the $z$ and $t$ 
directions, respectively, where $a$ is the lattice spacing and $L,L_z,L_t$ are
even integers. In the following, we will employ twisted boundary conditions
\cite{'tHooft:1979uj}
in much the same way as in
\cite{Luscher:1985wf,Snippe:1997ru}.
The twisted boundary conditions we use for gluons and quarks are
applied to the $(x,y)$ directions and are given by ($\nu=x,y$)
\begin{eqnarray}
U_\mu(x+L\hat{\nu}) & = & \Omega_\nu U_\mu(x) \Omega_\nu^{-1}\;, \\
\Psi(x+L\hat{\nu}) & = & \Omega_\nu \Psi(x) \Omega_\nu^{-1}\;,
\end{eqnarray}
where the quark field $\Psi_{sc}(x)$ becomes a matrix in smell-colour
space 
\cite{Parisi:1984cy}
by the introduction of a ``smell'' group SU($N_s$) with $N_s=N$ in
addition to the colour group SU($N$). We apply periodic boundary
conditions in the $(z,t)$ directions.

These boundary conditions lead to a change in the Fourier expansion of
the fields which now reads
\begin{eqnarray}
A_\mu(x) & = & \frac{1}{N L^2 L_z L_t} \sum^{\prime}_{p} \Gamma_p e^{ipx}
\tilde{A}_\mu(p)\;, \label{ft_glue}\\
\Psi_\alpha(x) & = & \frac{1}{N L^2 L_z L_t} \sum_{p} \Gamma_p e^{ipx}
\tilde{\Psi}_\alpha(x)\;.
\end{eqnarray}
In the twisted $(x,y)$ directions the sums are over
\begin{equation}
p_\nu = m n_\nu,~~-\frac{NL}{2} < n_\nu \le \frac{NL}{2},~~\nu = (x,y)\;,
\end{equation}
where $m = \frac{2\pi}{N L}$, and in the untwisted $(z,t)$ directions
the sums are over 
\begin{equation}
p_\nu = \frac{2\pi}{L_\nu} n_\nu,~~-\frac{L_\nu}{2} < n_\nu \le \frac{L_\nu}{2},~~\nu = (z,t)\;.
\end{equation}
The modes with ($n_x=n_y=0 \mod N$) are omitted from the sum in the case
of the gluons since the gauge group is non-abelian, and this is signified by
the prime on the summation symbol in Eqn. (\ref{ft_glue}). In particular, this
removes the zero mode from the gluon spectrum and so the mass-scale
$m$ defined above acts as a gauge-invariant infra-red regulator. The
matrices $\Gamma_p$ are given by  (up to an arbitrary phase, which may
be chosen for convenience)
\begin{equation}
\Gamma_p = \Omega_x^{-n_y}\Omega_y^{n_x}
\end{equation}
The momentum sums for quark loops need to be divided by $N$ to remove
the redundant smell factor.

The twisted theory can be viewed as a two-dimensional field theory in
the infinite $(z,t)$ plane with the modes in the twisted directions
being considered in the spirit of Kaluza-Klein modes. Denoting
$\bn=(n_x,n_y)$, the stable particles in the $(z,t)$ continuum limit
of this effective theory are called the A mesons ($\bn=(1,0)$ or
$\bn=(0,1)$) with mass $m$ and the B mesons ($\bn=(1,1)$) with mass
$\sqrt{2}m$ 
\cite{Snippe:1997ru}.
\subsection{Small-mass expansion}
\label{subsec:smallmass}

To extract the $\mathcal{O}(a^2)$ lattice artifacts, we first expand
some observable quantity $Q$ in powers of $ma$ at fixed $m_qa$:
\begin{eqnarray}\label{eqn:fit_in_ma}
Q(ma,m_qa)&=&a^{(Q)}_0(m_qa) + a^{(Q)}_2(m_qa) (ma)^2 + \nonumber\\
&&\mathcal{O}\left((ma)^4,(ma)^4\log(ma)\right)\label{eqn:Q}
\end{eqnarray}
where the coefficients in the expansion are all functions of $m_qa$.
There is no term at $\mathcal{O}\left((ma)^2\log(ma)\right)$ since the
gluon action is improved at tree-level to $O(a^2)$
\cite{Snippe:1997ru}.
Since we wish to extrapolate to the chiral limit it might be thought that 
we can set $m_qa=0$ straight away to achieve this end. However, the
correct chiral limit is $m_qa \to 0,~ma \to 0,~m_q/m > C$, where
$m = \frac{2\pi}{NL}$ as before and $C$ is a constant determined by
the requirement that the appropriate Wick rotation can be performed in
order to evaluate the Feynman integrals. If the inequality is violated
this results in a pinch singularity. It is physically sensible that
the correct limit is $L \to \infty$ before  $m_q \to 0$ since this
divorces the two infra-red scales and avoids complication. This does,
however, require us to consider the double expansion in $m_qa,ma$ and
carry out the extrapolation to $m_qa=0$ for the coefficients in
Eqn. (\ref{eqn:Q}). We return to this issue in the next section when
we discuss choice of integration contours. 

To extrapolate to the chiral limit, $m_qa \to 0$, we will fit the
coefficients in the expansion for $Q$ in $ma$ to their most general 
expansion in $m_qa$ for small $m_qa$. 

For $a^{(Q)}_0(m_qa)$ we have
\begin{equation}\label{eqn:fit0_in_mqa}
a^{(Q)}_0(m_qa)~=~b^{(Q)}_{0,0}\log(m_qa) + a^{(Q)}_{0,0}\;.
\end{equation}
Since we expect a well-defined continuum limit, $a^{(Q)}_0(m_qa)$
cannot contain any negative powers of $m_qa$ but, depending on the
quantity $Q$, it may contain logarithms; $b^{(Q)}_{0,0}$ is the
continuum anomalous dimension associated with $Q$ which is determined
by a continuum calculation. 

There can be no terms in $(m_qa)^{2n}, n>0$ since these are obviously
non-zero in the limit $ma \to 0$, and there is no counterterm in the
gluon action that can compensate for a scaling violation of this kind.

For $a^{(Q)}_2(m_qa)$ we find
{\bacol
\begin{eqnarray}\label{eqn:fit2_in_mqa}
&&a^{(Q)}_2(m_qa)~=~\frac{a^{(Q)}_{2,-2}}{(m_qa)^2} + a^{(Q)}_{2,0}\\
&&+ \left(a^{(Q)}_{2,2} + b^{(Q)}_{2,2}\log(m_qa)\right)(m_qa)^2 +
   \mathcal{O}\left((m_qa)^4\right)\;.\nonumber
\end{eqnarray}
}
After multiplication by $(ma)^2$ the $(m_qa)^{-2}$ contribution gives
rise to a continuum contribution to $Q$, and $a^{(Q)}_{2,-2}$ is
calculable in continuum perturbation theory. There can be no term in
$(m_qa)^{-2}\log(m_qa)$ since this would be a volume-dependent further
contribution to the anomalous dimension of $Q$, and there can be no
term in $\log(m_qa)$ since the action is tree-level $O(a^2)$ improved
\cite{Symanzik:1983dc}.

Depending on the choice of observable $Q$ there may be additional constraints
on the coefficients which appear in the expansions. We discuss these in the
next section in the context of the particular observables with which
we concern ourselves.

\section{Calculations and Results}
\label{sec:calculations}

In this section we lay out the calculation of the unquenching effects to 
order $\mathcal{O}(\alpha_s N_f a^2)$. The numbers and quantities given 
in the following are per quark flavour, and hence need to be multiplied 
by $N_f$ throughout.

\subsection{The A meson mass}
\label{subsec:amass}

\begin{figure}
\includegraphics[width=\myfigwidth,keepaspectratio=,clip=]{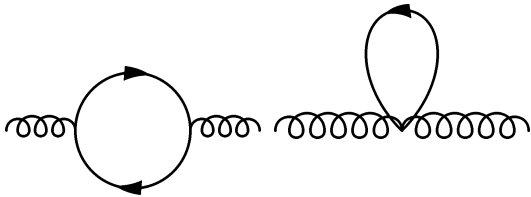}
\caption{The fermionic one-loop diagrams contributing to the A meson
  mass renormalisation as well as to the wavefunction renormalisation
  for A and B mesons}
\label{fig:amass_diagrams}
\end{figure}

The simplest spectral quantity that can be chosen within the framework
of the twisted boundary conditions outlined above is the
(renormalised) mass of the A meson. In agreement with Eqn. (109) of
\cite{Snippe:1997ru}
the one-loop correction the the A meson mass (for A mesons with
positive spin) is given by
\begin{equation}\label{eqn:mA}
m_A^{(1)} = - Z_0(\mathbf{k})
              \left.\frac{\pi_{11}^{(1)}(k)}{2 m_A^{(0)}}\right|
              _{k=(i m_A^{(0)},0,m,0)}
\end{equation}
where $Z_0(\mathbf{k})=1+\mathcal{O}\left((ma)^4\right)$ is the
residue of the pole of the tree-level gluon propagator at spatial
momentum $\mathbf{k}$, and $m_A^{(0)}$ is defined so that the momentum
$k$ is on-shell. We consider the dimensionless quantity $m_A^{(1)}/m$.
The fermionic diagrams that contribute to this quantity are shown in
figure
\ref{fig:amass_diagrams}.

The anomalous dimension of $m_A$ is zero and so using 
Eqn. (\ref{eqn:fit0_in_mqa}) we have 
\begin{equation}\label{eqn:anom_dir_m_A}
b^{(m_A,1)}_{0,0}~=~0\;.
\end{equation}
Using physical arguments we can determine the behaviour of other
coefficients. From continuum calculations we find
\begin{equation}
a^{(m_A,1)}_{2,-2}~=~0\;.
\end{equation}
This result follows from the fact that the fermion
contribution at one loop order to $m_A$ is IR finite since the fermion
has a non-zero mass and is 4D Lorentz invariant. Thus,
$a^{(m_A)}_{2,-2}$ can be constructed only from 4D Lorentz invariants
of which we have only $\be_A\cdot \bk_A$ and $\bk_A^2$, where $\be_A$
is the $A$-meson polarization vector. However, gauge invariance
implies $\be_A\cdot \bk_A=0$ and the on-shell condition gives
$\bk_A^2=0$ and so, there being no non-zero Lorentz invariant, we
deduce the result.  

A much less obvious deduction is that $a^{(m_A,1)}_0(m_qa)=0$, which
together with eqns.
(\ref{eqn:fit0_in_mqa},\ref{eqn:anom_dir_m_A})
implies that $a^{(m_A,1)}_{0,0}=0$. A necessary ingredient to derive
this result is the fact that the one-loop fermion contribution to
$m_A$ is IR finite in the limit $m \to 0$ ($L \to \infty$) since the
fermion mass, $m_q$, is non-zero. We thus expect that in this limit
Lorentz invariance will be restored, although for $L$ finite this will
not be the case. Gauge invariance and the Ward Identity then ensure
that, in this limit, Lorentz invariance implies that the gluon
self-energy function $\pi_{\mu\nu}(\bk)$ satisfies
\begin{equation}\label{eqn:gluon_pi}
\pi_{\mu\nu}(\bk)~=~(k^2g_{\mu\nu}-k_\mu k_\nu)\pi(k^2)\;.
\end{equation}
From
\cite{Luscher:1985wf,Snippe:1997ru}
the one-loop contribution to $m_A$ is proportional
to $\be^\mu\be^\nu\pi_{\mu\nu}(\bk)$. In the limit $L \to \infty$ we
are able to use Eqn.
(\ref{eqn:gluon_pi})
and we find that this contribution is zero by gauge invariance and the
on-shell condition $k^2=0$. In contrast, the contribution to
$a^{(m_A)}_0$ from internal gluon loops is not zero by this argument;
it is indeed calculated in
\cite{Luscher:1985wf,Snippe:1997ru}.
The reason is that the one-loop gluon contribution is not IR finite in
the $m \to 0,~L \to \infty$ limit since the IR-regulating mass is $m$;
the internal gluon ``feels'' the finite boundary of the lattice in the
$x,y$ direction no matter how large $L$ is. Consequently, we cannot
expect to use restoration of Lorentz invariance to limit the form that
the purely gluonic $\pi_{\mu\nu}(\bk)$ takes, and so no deduction
concerning this coefficient can be made.

An alternative explanation for why $a^{(m_A,1)}_0=0$ also relies on
the restoration of Lorentz invariance in the $m \to 0$ limit. In this
limit, the action is isotropic with metric tensor $g_{\mu\nu} =
\mbox{diag}(1,1,1,1)$. However, the twisted boundary conditions break
Lorentz invariance and single out the twisted $x,y$ directions, and so
we must expect that radiative corrections will renormalise
$g_{\mu\nu}$ in a way that can break Lorentz symmetry. The mass shell
condition for the $A$-meson is then
\begin{equation}
g^R_{\mu\nu}k^\mu k^\nu=0,~~~k_\mu = (ip_0,0,m,k_3)\;,
\end{equation}
where $g^R$ is the renormalised metric tensor. This is reinterpreted
as a renormalization of the $A$-meson mass $m$ with
\begin{equation}
m^R_A~=~\frac{g^R_{11}(m)}{g_{11}}m\;.
\end{equation}
This can also be interpreted as an anisotropy renormalization. Since
the one-loop fermion contribution is IR finite and Lorentz symmetry is
restored in the limit $m \to 0$ we then have that
$g^R_{11}(m=0)=g_{11}=1$ and $m_A$ is not renormalised. This is not
the case for the one-loop gluon contribution, which is not IR finite,
and so the assumption that Lorentz symmetry is restored as $m \to 0$
is incorrect.

\begin{figure}[b]
\includegraphics[height=\myfigwidth,angle=90,keepaspectratio=,clip=]{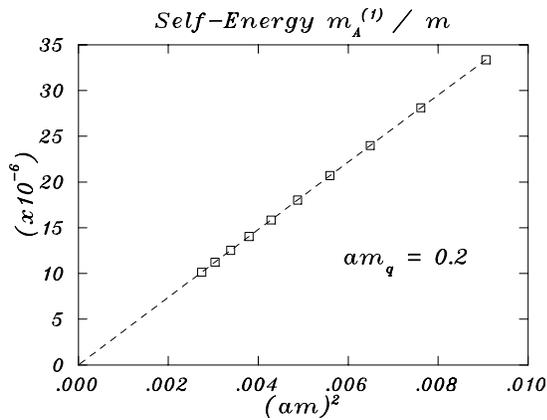}
\caption{A plot of the fermionic contributions to the one-loop $A$
  meson self-energy $m_A^{(1)}/m$ against $(ma)^2$. The vanishing of
  $m_A^{(1)}/m$ in the infinite-volume limit can be seen clearly.}
\label{fig:mA_vs_ma2}
\end{figure}

For the kinematics used here this means that in the limit
$m \to 0,~L \to \infty$ then $\pi_{11}$ vanishes and hence from
Eqn. (\ref{eqn:mA}) so does $m_A^{(1)}/m$. This expectation is
accurately verified by our calculation: the extrapolation of
$m_A^{(1)}/m$ to $m = 0$ indeed gives zero (cf. figure \ref{fig:mA_vs_ma2}).

In the chiral limit $m_q\to 0$, the term $w_i$ that appears on the
right-hand side of Eqn.
(\ref{eqn:impcond_generic})
is $a^{(Q)}_{2,0}$, and it is this limit and this coefficient that we
will concern ourselves with hereafter.

The $\mathcal{O}\left(\alpha_s (ma)^2\right)$ contribution from
improvement of the action is given by
\cite{Snippe:1997ru}
\begin{equation}
\Delta_\textrm{imp} \frac{m_A^{(1)}}{m} = 
- ( c_1^{(1)} - c_2^{(1)} ) (ma)^2 + \mathcal{O}\left((ma)^4\right)
\end{equation}
leading to the improvement condition
\begin{equation}
\label{eqn:impcond_mA}
c_1^{(1)} - c_2^{(1)} = a_{2,0}^{(m_A,1)}
\end{equation}

\subsection{The three-point coupling}
\label{subsec:threept}

\begin{figure}
\includegraphics[width=\myfigwidth,keepaspectratio=,clip=]{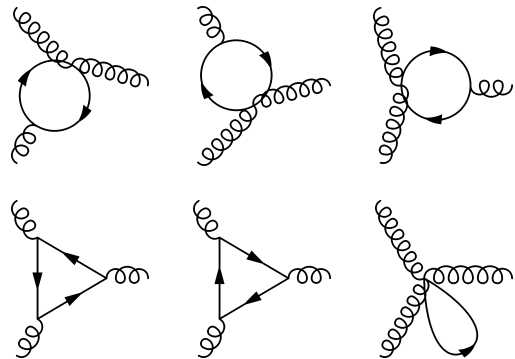}
\caption{The fermionic one-loop diagrams contributing to the
  three-point function.}
\label{fig:threept_diagrams}
\end{figure}

An effective coupling constant $\lambda$ for an AAB meson vertex is
defined as
\begin{equation}
\label{eqn:def_of_lambda}
\lambda =
g_0 \sqrt{ Z(\mathbf{k}) Z(\mathbf{p}) Z(\mathbf{q}) }
e_j \Gamma^{1,2,j}(k,p,q)
\end{equation}
where we have factored out a twist factor of
$\frac{i}{N}\Tr([\Gamma_k,\Gamma_p]\Gamma_q)$ from both sides, and the
momenta and polarisations of the incoming particles are
\begin{eqnarray}\label{eqn:momenta}
k = (iE(\mathbf{k}),\mathbf{k});&& \mathbf{k} = (0,m,ir)\nonumber\\
p = (-iE(\mathbf{p}),\mathbf{p});&& \mathbf{p} = (m,0,ir)\nonumber\\
q = (0,\mathbf{q});&& \mathbf{q} = (-m,-m,-2ir)\nonumber\\
e = (0,1,-1,0) &&
\end{eqnarray}
Here $r>0$ is defined such that $E(\mathbf{q})=0$. This coupling is a
spectral quantity since it can be related to the scattering amplitude
of A mesons
\cite{Luscher:1985zq}.
We expand Eqn.
(\ref{eqn:def_of_lambda})
perturbatively to one-loop order and find in agreement with Eqn. (137)
of
\cite{Snippe:1997ru}
\begin{eqnarray}
\frac{\lambda^{(1)}}{m} & = &
\left( 1 - \smallfrac{1}{24} m^2 \right)\frac{\Gamma^{(1)}}{m}
- \frac{4}{k_0} \frac{d}{dk_0}
           \left.\pi_{11}^{(1)}(k)\right|_{k_0=iE(\mathbf{k}}) \\
&& - \left( 1 - \smallfrac{1}{12} m^2 \right) \frac{d^2}{dq_0^2}
           \left. \left( e^i e^j \pi_{ij}^{(1)}(q) \right) \right|_{q_0=0}
+ \mathcal{O}(m^4) \nonumber
\end{eqnarray}
The fermionic diagrams contributing to the irreducible three-point
function $\Gamma^{(1)}$ are shown in figure
\ref{fig:threept_diagrams}.
Using Eqn. (\ref{eqn:fit0_in_mqa}) and the known anomalous dimension
of the coupling constant we have
\begin{equation}\label{eqn:anom_dir_lambda}
b^{(\lambda,1)}_{0,0}~=~-\frac{N_f}{3\pi^2}g^2\;.
\end{equation}
Unlike the argument for $a^{(m_A,1)}_{2,-2}~=~0$ above, a continuum
calculation gives
\begin{equation}
a^{(\lambda,1)}_{2,-2}~=~-\frac{N_f}{120\pi^2}g^2\;.
\end{equation} 
In this case there are non-zero Lorentz invariants for the three-point 
function such as $\be_A\cdot \bk_B$ etc. and so we expect this coefficient 
to be non-zero.

The improvement contribution to $\lambda$ is
\cite{Snippe:1997ru}
\begin{equation}
\Delta_\textrm{imp} \frac{\lambda^{1}}{m} = 4 (9 c_1^{(1)} - 7 c_2^{(1)}) (ma)^2
   + \mathcal{O}\left((ma)^4\right)
\end{equation}
leading to the improvement condition
\begin{equation}
\label{eqn:impcond_lambda}
4 (9 c_1^{(1)} - 7 c_2^{(1)}) = - a_{2,0}^{(\lambda,1)}
\end{equation}

\begin{figure}[t]
\includegraphics[height=\myfigwidth,angle=90,keepaspectratio=,clip=]{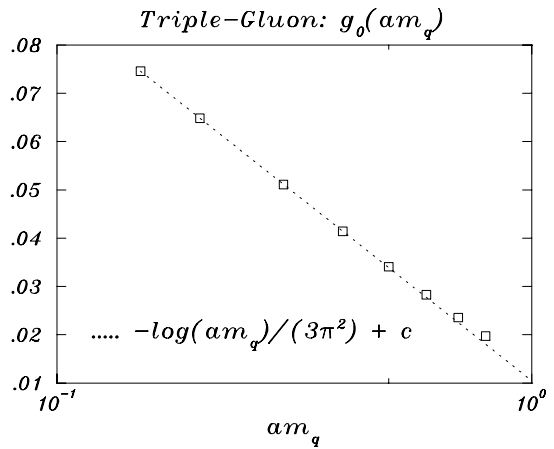}
\caption{A plot of $a_0^{(\lambda,1)}$ against $m_qa$ which shows the
  agreement between the numerical lattice results and the known
  anomalous dimension.}
\label{fig:a0l_vs_mqa}
\end{figure}

\begin{figure}[b]
\includegraphics[height=\myfigwidth,angle=90,keepaspectratio=,clip=]{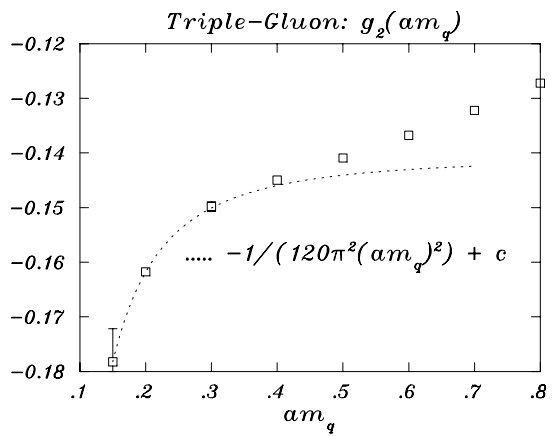}
\caption{A plot of $a_2^{(\lambda,1)}$ against $m_qa$ with the
  analytical continuum result for the infrared divergence shown for
  comparison.}
\label{fig:a2l_vs_mqa}
\end{figure}

Tests of our calculation are that the fit for $a^{(\lambda,1)}_{0,0}$
must give the correct anomalous dimension stated in Eqn.
(\ref{eqn:anom_dir_lambda}),
and that our fits reproduce the continuum
result $a^{(\lambda,1)}_{2,-2}=-g^2/120\pi$. Both are accurately
verified (cf. figures \ref{fig:a0l_vs_mqa} and \ref{fig:a2l_vs_mqa}).

\subsection{Choice of integration contours}
\label{subsec:contours}

The external lines of the diagrams are on their respective mass shells
but with complex three-momentum $\bk$ in which the third component,
$k_3$, has been continued to an imaginary value parametrised by the
variable $r$ as shown in Eqn.
(\ref{eqn:momenta});
in the Euclidean formulation $k_0$ is also imaginary. In evaluating
the loop integrals that are not pure tadpoles, care must be taken to
ensure that the amplitudes calculated are the correct analytic
continuations in $r$ from the Minkowski space on-shell amplitudes
defined with real three-momenta to the ones in
Eqn.
(\ref{eqn:momenta}).

The situation is complicated by the presence of two mass scales $m,
m_q$. The integrals are evaluated after performing a Wick rotation in
$k_0$, taking care to avoid  contour crossing of any poles that move
as $r$ is continued from $r=0$ to $r=m/\sqrt{2}$. We find that $m_q/m$
must be chosen larger than a minimum value, dependent on the graph
being considered, to avoid any contour being pinched. The outcome is
that after the Wick rotation in $k_0$, the (Euclidean) integration
contour for either $k_0$ or, in one case, $k_3$ must be shifted by an
imaginary constant.

We find it sufficient that for the calculation of $m_A^{(1)}$ and
$Z_A(k)$ we impose $m_q > m/2$ and for $\Gamma^{(1)}$ and $Z_B(q)$
that $m_q > m/\sqrt{2}$.

\subsection{Extracting the coefficients}
\label{subsec:fitting}

To extract the improvement coefficients from our diagrammatic
calculations, we compute the diagrams for a number of different values
of both $L$ and $m_q$ with $N_f=1$, $N=3$. At each value of $m_q$, we
then perform a fit in $ma$ of the form given in Eqn.
(\ref{eqn:fit_in_ma})
to extract the coefficients $a_n^{(Q,1)}(m_qa),~n=0,2$. The results of these
fits are given in Table
\ref{tab:fitting_coeffs}. 
\begin{table}
\begin{tabular}{llll}\hline\hline
$m_qa$&$a_2^{(m_A,1)}$   &$a_0^{(\lambda,1)}$&$a_2^{(\lambda,1)}$\\\hline
0.15  & 0.0036752(7)    & 0.07456(1)         & -0.178(6)           \\
0.2   & 0.003701(1)     & 0.0648161(5)       & -0.1617(4)          \\
0.3   & 0.003730711(1)  & 0.051090(2)        & -0.1498(9)          \\
0.4   & 0.00372996(4)   & 0.0414332(1)       & -0.14498(6)         \\
0.5   & 0.003696507(2)  & 0.03408572(2)      & -0.140933(7)        \\
0.6   & 0.0036328671(4) & 0.028272563(9)     & -0.136776(2)        \\
0.7   & 0.0035435429(3) & 0.023575013(4)     & -0.132222(1)        \\
0.8   & 0.0034337690(4) & 0.019733513(1)     & -0.12722(3)         \\
0.9   & 0.0033087971(2) & ---                & ---                 \\
1.0   & 0.0031734700(3) & ---                & ---                 \\
1.2   & 0.0028882(3)    & 0.009976(2)        & -0.1044(1)          \\
\hline\hline
\end{tabular}
\caption{The coefficients from the fits of $m_A^{(1)}$ and
$\lambda^{(1)}/m$ against $ma$.}
\label{tab:fitting_coeffs}
\end{table}

To facilitate our fits, we make use of the prior physical
knowledge we have: In the case of $m_A^{(1)}$, we have
$a_0^{(m_A,1)}=0$ because of gauge invariance.

Performing a fit of the form
(\ref{eqn:fit0_in_mqa}) and (\ref{eqn:fit2_in_mqa}),
respectively, on these coefficients, we get the required coefficients 
of the $\mathcal{O}(a^2)$ lattice artifacts in the chiral limit to be
\begin{eqnarray}
a_{2,0}^{(m_A,1)} & = & 0.00361(1) \\
a_{2,0}^{(\lambda,1)} & = & -0.140(1)
\end{eqnarray}
These coefficients are to be identified with the $w_i$ of Eqn.
(\ref{eqn:impcond_generic}).

Here, again, we have facilitated our fits by making use of our prior
knowledge: For $m_A^{(1)}$, $a_{2,-2}^{(m_A,1)}$  vanishes, and for
$\lambda^{(1)}$, we have two known continuum contributions:
$b_{0,0}^{(\lambda,1)}=-1/3\pi^2$ is the one-loop coefficient
of the $\beta$-function and
$a_{2,-2}^{(\lambda,1)}=-1/120\pi^2$ is the continuum
coefficient of the infrared divergence $m^2/m_q^2$.

Solving eqns.
(\ref{eqn:impcond_mA}) and (\ref{eqn:impcond_lambda})
for $c_i^{(1)}$, our results can be summarised as
\begin{eqnarray}
c_1^{(1)} & = & -0.025218(4) + 0.00486(13) N_f \\
c_2^{(1)} & = & -0.004418(4) + 0.00126(13) N_f
\end{eqnarray}
where the quenched ($N_f=0$) results are taken from
\cite{Snippe:1997ru}.
\section{Conclusions}

Repeating the analysis of
\cite{Alford:1995hw}
and using their notation we express the radiatively corrected action
of Eqn.
(\ref{eqn:lw_action})
as
\cite{Luscher:1984xn,Alford:1995hw}
\begin{equation}
S[U]~=~\sum_{i=0}^2 \beta_iS_i[U]
\end{equation}
Then 
\begin{eqnarray}
\beta_1&=&-\frac{\beta_0}{20u_0^2}\left[1-\left(\frac{12\pi}{5}c_0^{(1)}+
 48\pi c_1^{(1)}+2u_0^{(1)}\right)\alpha_s\right]\;,\nonumber\\
\beta_2&=&\frac{12\pi\beta_0}{5u_0^2}c_2^{(1)}\alpha_s\;.
\end{eqnarray}
The quenched radiative contributions have been analyzed in
\cite{Alford:1995hw}
and so we may write
\begin{eqnarray}
\beta_1&=&-\frac{\beta_0}{20u_0^2}\left[1+0.4805\alpha_s-\left(\frac{12\pi}{5}c_{0,f}^{(1)}+
48\pi c_{1,f}^{(1)}\right)\alpha_s\right]\;,\nonumber\\
\beta_2&=&-\frac{\beta_0}{u_0^2}\left(0.033\alpha_s-\frac{12\pi}{5}c_{2,f}^{(1)}\alpha_s\right)\;,
\end{eqnarray}
where now all the one-loop coefficients $c_{i,f}^{(1)}$ contain only
quark loop contributions.

Plugging in the numbers obtained in this work we find
\begin{eqnarray}
\beta_1&=&-\frac{\beta_0}{20u_0^2}\left[1+0.4805\alpha_s-0.3637(14)N_f\alpha_s\right]\;,\nonumber\\
\beta_2&=&-\frac{\beta_0}{u_0^2}\left(0.033\alpha_s-0.009(1)N_f\alpha_s\right)\;.
\end{eqnarray}
With $N_f=3$ the shift from the quenched values is surprisingly large,
and may have a significant impact; especially, it may explain the
increased scaling violations seen in some unquenched simulations.

\section*{Acknowledgments}

The authors thank Fermilab for the use of computing resources.
This work was supported in part by
%% GMvH is partially supported by
the Canadian Natural Sciences and Engineering Research Council
(NSERC), the Government of Saskatchewan,
%% QJM was partially supported by
and the UK Particle Physics and Astronomy Research Council (PPARC).

%
%
%
%
%%\bibliographystyle{h-physrev4}
%%\bibliography{nf_lw_imp}

\begin{thebibliography}{10}

\bibitem{Davies:2003ik}
C.~T.~H. Davies {\em et~al.},
\newblock Phys. Rev. Lett. {\bf 92}, 022001 (2004), [hep-lat/0304004];
%%CITATION = HEP-LAT 0304004;%%

\bibitem{Aubin:2004fs}
C. Aubin {\em et~al.},
\newblock Phys. Rev. {\bf D70}, 114501 (2004), [hep-lat/0407028].
%%CITATION = HEP-LAT 0407028;%%

\bibitem{Mason:2005zx}
Q.~Mason {\em et~al.},
\newblock Phys. Rev. Lett. {\bf 95}, 052002 (2005), [hep-lat/0503005].
%%CITATION = HEP-LAT 0503005;%%

\bibitem{Mason:2005bj}
Q.~Mason, H.~D. Trottier, R.~Horgan, C.~T.~H. Davies and G.~P. Lepage,
\newblock Phys. Rev. {\bf D73}, 114501 (2006), [hep-ph/0511160].
%%CITATION = HEP-PH 0511160;%%

\bibitem{Kronfeld:2005fy}
A.~S. Kronfeld {\em et~al.},
\newblock PoS {\bf LAT2005}, 206 (2006), [hep-lat/0509169].
%%CITATION = HEP-LAT 0509169;%%

\bibitem{Lepage:1998vj}
G.~P. Lepage,
\newblock Phys. Rev. {\bf D59}, 074502 (1999), [hep-lat/9809157].
%%CITATION = HEP-LAT 9809157;%%

\bibitem{Orginos:1999cr}
K.~Orginos, D.~Toussaint and R.~L. Sugar,
\newblock Phys. Rev. {\bf D60}, 054503 (1999), [hep-lat/9903032].
%%CITATION = HEP-LAT 9903032;%%

\bibitem{Sharpe:2006re}
S.~R.~Sharpe,
\newblock PoS {\bf LAT2006}, 022 (2006), [hep-lat/0610094].
%%CITATION = POSCI,LAT2006,022;%%

\bibitem{Davies:private}
C.~T.~H. Davies,
\newblock private communication.

\bibitem{Luscher:1985wf}
M.~L\"uscher and P.~Weisz,
\newblock Nucl. Phys. {\bf B266}, 309 (1986).
%%CITATION = NUPHA,B266,309;%%

\bibitem{Snippe:1997ru}
J.~Snippe,
\newblock Nucl. Phys. {\bf B498}, 347 (1997), [hep-lat/9701002].
%%CITATION = HEP-LAT 9701002;%%

\bibitem{Hao:2006}
Zh. Hao,
\newblock M.Sc. thesis, Simon Fraser University 2006.

\bibitem{Symanzik:1983dc}
K.~Symanzik,
\newblock Nucl. Phys. {\bf B226}, 187 (1983).
%%CITATION = NUPHA,B226,187;%%

\bibitem{Symanzik:1983gh}
K.~Symanzik,
\newblock Nucl. Phys. {\bf B226}, 205 (1983).
%%CITATION = NUPHA,B226,205;%%

\bibitem{Luscher:1984xn}
M.~L\"uscher and P.~Weisz,
\newblock Commun. Math. Phys. {\bf 97}, 59 (1985).
%%CITATION = CMPHA,97,59;%%

\bibitem{Luscher:1985zq}
M.~L\"uscher and P.~Weisz,
\newblock Phys. Lett. {\bf B158}, 250 (1985).
%%CITATION = PHLTA,B158,250;%%

\bibitem{Alford:1995hw}
M.~G. Alford, W.~Dimm, G.~P. Lepage, G.~Hockney and P.~B. Mackenzie,
\newblock Phys. Lett. {\bf B361}, 87 (1995), [hep-lat/9507010].
%%CITATION = HEP-LAT 9507010;%%

\bibitem{Drummond:2002kp}
I.~T. Drummond, A.~Hart, R.~R. Horgan and L.~C. Storoni,
\newblock Nucl. Phys. Proc. Suppl. {\bf 119}, 470 (2003), [hep-lat/0209130].
%%CITATION = HEP-LAT 0209130;%%

\bibitem{Hart:2004bd}
A.~Hart, G.~M. von Hippel, R.~R. Horgan and L.~C. Storoni,
\newblock J. Comput. Phys. {\bf 209}, 340 (2005), [hep-lat/0411026].
%%CITATION = HEP-LAT 0411026;%%

\bibitem{Nobes:2001tf}
M.~A. Nobes, H.~D. Trottier, G.~P. Lepage and Q.~Mason,
\newblock Nucl. Phys. Proc. Suppl. {\bf 106}, 838 (2002), [hep-lat/0110051].
%%CITATION = HEP-LAT 0110051;%%

\bibitem{Nobes:2003nc}
M.~A. Nobes and H.~D. Trottier,
\newblock Nucl. Phys. Proc. Suppl. {\bf 129}, 355 (2004), [hep-lat/0309086].
%%CITATION = HEP-LAT 0309086;%%

\bibitem{Trottier:2003bw}
H.~D. Trottier,
\newblock Nucl. Phys. Proc. Suppl. {\bf 129}, 142 (2004), [hep-lat/0310044].
%%CITATION = HEP-LAT 0310044;%%

\bibitem{'tHooft:1979uj}
G.~'t~Hooft,
\newblock Nucl. Phys. {\bf B153}, 141 (1979).
%%CITATION = NUPHA,B153,141;%%

\bibitem{Parisi:1984cy}
G.~Parisi,
\newblock Invited talk given at Summer Inst. Progress in Gauge Field Theory,
  Cargese, France, Sep 1-15, 1983.

\end{thebibliography}

\end{document}